\documentclass[aip,jcp,notitlepage,twocolumn,reprint,floatfix]{revtex4-2}

\usepackage{graphicx}
\usepackage{dcolumn}
\usepackage{bm}
\usepackage{hyperref}
\usepackage{amsmath}
\usepackage{listings}
\usepackage{mathtools}
\usepackage{booktabs}
\usepackage{multirow}
\usepackage{comment}
\usepackage{natbib}
\usepackage{xcolor}
\let\oldtheequation\theequation
\makeatletter
\def\tagform@#1{\maketag@@@{\ignorespaces#1\unskip\@@italiccorr}}
\renewcommand{\theequation}{(\oldtheequation)}
\makeatother

\newcommand{\appref}[1]{\hyperref[#1]{Appendix~\ref{#1}}}
\newcommand{\lit}[1]{Ref.~\onlinecite{#1}}

\definecolor{codegreen}{rgb}{0.3,0.65,0.1}
\definecolor{codegray}{rgb}{0.5,0.5,0.5}
\definecolor{codepurple}{rgb}{0.6,0.3,0.6}
\definecolor{codeblue}{rgb}{0.3,0.6,0.95}
\definecolor{backcolor}{rgb}{0.95,0.95,0.95}

\begin{document}

\title{\textsc{Block2}: a comprehensive open source framework to develop and apply state-of-the-art DMRG algorithms in electronic structure and beyond}

\author{Huanchen Zhai}
\email{hczhai.ok@gmail.com}
\affiliation{Division of Chemistry and Chemical Engineering, California Institute of Technology, Pasadena, CA 91125, USA}

\author{Henrik R.~Larsson}
\thanks{Present address: Department of Chemistry and Biochemistry, University of California, Merced, CA 95343, USA}
\affiliation{Division of Chemistry and Chemical Engineering, California Institute of Technology, Pasadena, CA 91125, USA}

\author{Seunghoon Lee}
\thanks{Present address: Department of Chemistry, Seoul National University, Seoul 151-747, South Korea}
\affiliation{Division of Chemistry and Chemical Engineering, California Institute of Technology, Pasadena, CA 91125, USA}

\author{Zhi-Hao Cui}
\thanks{Present address: Department of Chemistry, Columbia University, New York, NY 10027, USA}
\affiliation{Division of Chemistry and Chemical Engineering, California Institute of Technology, Pasadena, CA 91125, USA}

\author{Tianyu Zhu}
\thanks{Present address: Department of Chemistry, Yale University, New Haven, CT 06520, USA}
\affiliation{Division of Chemistry and Chemical Engineering, California Institute of Technology, Pasadena, CA 91125, USA}

\author{Chong Sun}
\thanks{Present address: Department of Chemistry, Rice University, Houston, Texas 77005, USA}
\affiliation{Division of Chemistry and Chemical Engineering, California Institute of Technology, Pasadena, CA 91125, USA}

\author{Linqing Peng}
\affiliation{Division of Chemistry and Chemical Engineering, California Institute of Technology, Pasadena, CA 91125, USA}

\author{Ruojing Peng}
\affiliation{Division of Chemistry and Chemical Engineering, California Institute of Technology, Pasadena, CA 91125, USA}

\author{Ke Liao}
\thanks{Present address: Department of Physics and Arnold Sommerfeld Center for Theoretical Physics, Ludwig-Maximilians-Universit\"{a}t M\"{u}nchen, D-80333 Munich, Germany}
\affiliation{Division of Chemistry and Chemical Engineering, California Institute of Technology, Pasadena, CA 91125, USA}

\author{Johannes T{\"o}lle}
\affiliation{Division of Chemistry and Chemical Engineering, California Institute of Technology, Pasadena, CA 91125, USA}

\author{Junjie Yang}
\affiliation{Division of Chemistry and Chemical Engineering, California Institute of Technology, Pasadena, CA 91125, USA}

\author{Shuoxue Li}
\affiliation{Division of Chemistry and Chemical Engineering, California Institute of Technology, Pasadena, CA 91125, USA}

\author{Garnet Kin-Lic Chan}
\email{gkc1000@gmail.com}
\affiliation{Division of Chemistry and Chemical Engineering, California Institute of Technology, Pasadena, CA 91125, USA}

\date{\today}

\begin{abstract}
\textsc{block2} is an open source framework to implement and perform density matrix renormalization group and matrix product state algorithms. Out-of-the-box it supports the eigenstate,  time-dependent, response, and finite-temperature algorithms. In addition, it carries special optimizations for \emph{ab initio} electronic structure Hamiltonians and implements many quantum chemistry extensions to the density matrix renormalization group, such as dynamical correlation theories. The code is designed with an emphasis on flexibility, extensibility, and efficiency, and to support integration with external numerical packages. Here we explain the design principles and currently supported features and present numerical examples in a range of applications.
\end{abstract}

\maketitle

\section{Introduction}

The density matrix renormalization group (DMRG) algorithm\cite{white1992density,white1993density,verstraete2023density} is now a widely used numerical approach to obtain the low-energy eigenstates and other properties of a wide range of systems of interest in electronic structure, such as strongly correlated low-dimensional models,\cite{hachmann2006multireference,singh2010simulation,leblanc2015solutions,motta2017towards,motta2020ground} large active space problems in quantum chemistry,\cite{hachmann2007radical,marti2008density,kurashige2009high,mizukami2010communication,kurashige2013entangled,harris2014ab,sharma2014low,chalupsky2014reactivity,li2019pcluster,larsson2022chromium} electronic quantum dynamics,\cite{frahm2019ultrafast,baiardi2021electron} and even models to benchmark quantum computing protocols.\cite{li2019electronic,lee2023evaluating} Over the last several decades, there have been many developments and extensions of the original algorithm for these applications.\cite{chan2011density,wouters2014density,baiardi2020density} For example, dynamical electron correlation,\cite{yanai2015density,cheng2022post} the effects of finite temperature,\cite{feiguin2005finite,stoudenmire2010minimally} relativistic effects,\cite{roemelt2015spin} time-dependent phenomena,\cite{feiguin2005time} and frequency-dependent\cite{jeckelmann2002dynamical} response properties, can now all be treated within the DMRG framework.

In parallel with these theoretical advances, we have also witnessed the success of many different DMRG implementations for electronic structure,\cite{fishman2022itensor,chan2002highly,sharma2012spin,legeza2003controlling,boguslawski2013orbital,kurashige2009high,kurashige2013entangled,luo2010optimizing,wouters2014chemps2,keller2015efficient,keller2016spin,li2017spin,brabec2021massively,xie2023kylin} which together have helped demonstrate the power of DMRG in different settings. However, as more extensions of the DMRG algorithm have appeared and as DMRG techniques have grown to encompass more applications, so has the complexity grown in many implementations.

The new open source \textsc{block2} code\cite{block2} introduced here, represents an effort 
to provide users with a comprehensive set of state-of-the-art DMRG algorithms used today in electronic structure and other applications, while enabling access to the development and extensions of these methods.
To this end, the core design principles in  \textsc{block2} aim to balance efficiency, flexibility, and simplicity.
For example:

(i) For efficiency and flexibility, we provide a special highly optimized DMRG implementation for the standard quantum chemistry Hamiltonian, but many of the optimizations also accelerate computations with arbitrary sparse Hamiltonians, thus enabling production level applications involving custom Hamiltonians.

(ii) For efficiency and simplicity, we provide  both optimized and unoptimized implementations of most features in the code, so a developer can always start with the simplest version to explore new ideas, or compare different implementations to understand our optimization techniques.
The framework supports both the matrix product state (MPS)/matrix product operator (MPO) and renormalized operator views of DMRG algorithms, balancing the ease of expressing algorithms with optimized performance.

(iii) For flexibility and simplicity, while the algorithms in \textsc{block2} support general Hamiltonian and operator expressions, e.g.~arbitrary-order interactions, arbitrary-order density matrix computations, and arbitrary-order excitations, the computations can be specified through a simple symbolic Python syntax.

We note that to achieve the above design goals, many implementations in \textsc{block2} use techniques that to our knowledge have not been published in the literature. The following does not attempt to provide a full description of the implementation, which after all comprises approximately 200,000 lines of code. Instead, we introduce a sample of the existing features of \textsc{block2}, some ideas behind the implementation, and how the capabilities are used in applications and algorithm development. 

\section{Program Features}

\subsection{Standard DMRG}

\textsc{block2} provides a highly optimized and flexible implementation of the DMRG algorithm for solving various aspects of the electronic structure problem using \emph{ab initio} and model Hamiltonians. This section presents an overview of the most common features of the code used in electronic structure calculations.

\subsubsection{\emph{Ab initio} DMRG and symmetries}

The most common application of the \emph{ab initio} DMRG algorithm is to compute a many-body ground state wavefunction with some desired symmetries.\cite{white1993density,zgid2008spin,sharma2012spin,sharma2015general,keller2016spin,li2017spin,li2023time} One first requires the Hamiltonian expressed in a basis of orthogonal orbitals, which form the sites in DMRG. Such orbitals, and the corresponding Hamiltonian integrals, can be computed by many quantum chemistry packages, for example, \textsc{PySCF}. \cite{sun2018pyscf,sun2020recent}

The Hamiltonian expressed in the chosen orbital basis typically has a number of symmetries.
\textsc{block2} has built-in efficient support for many different symmetries [including particle-number U(1),\cite{chan2002highly} total spin SU(2),\cite{sharma2012spin} projected spin U(1),\cite{chan2002highly}  \(K\)-point \(\mathrm{Z}_n\),\cite{xiang1996density} SO(4) in the Hubbard model,\cite{zhang1991so,mcculloch2002non} and the Abelian point group (\(\mathrm{D_{2h}}\) and its subgroups\cite{wouters2014density})] 
and is designed to be easily extensible to other symmetries.

\begin{figure}[!htbp]
  \includegraphics[width=\columnwidth]{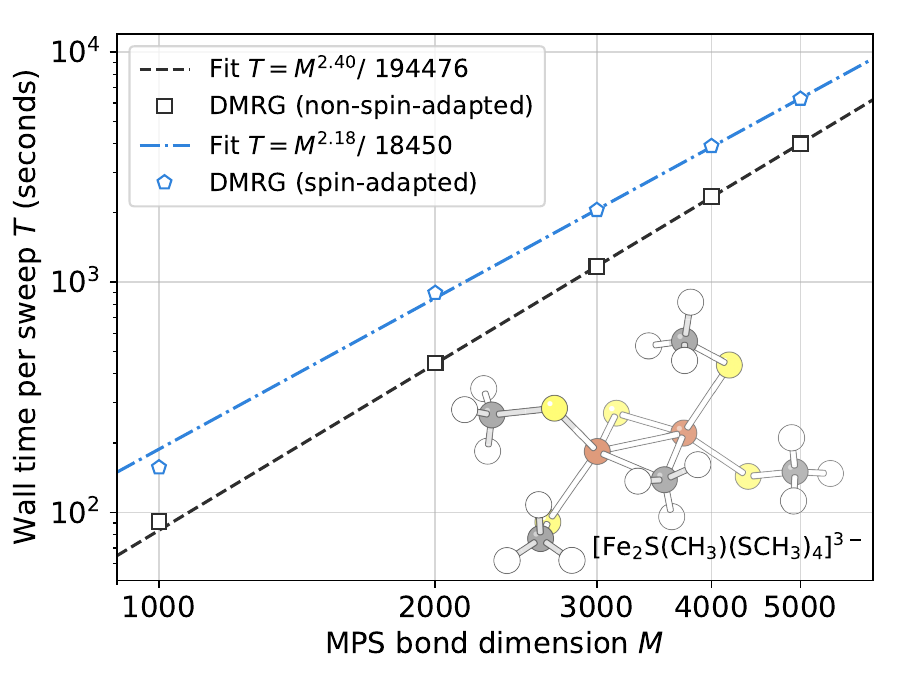}
  \caption{The scaling of DMRG wall time (using 24 CPU cores) per sweep with respect to the MPS bond dimension \(M\), for the (36o, 48e) active space of a protonated iron sulfur dimer \( [\mathrm{Fe_2S(CH_3)(SCH_3)_4}]^{3-} \). The active space model is from \lit{zhai2023multireference}. Spin-adapted uses particle-number and SU(2) symmetry, while non-spin-adapted uses particle-number and projected-spin U(1) symmetry.}
  \label{fig:symm}
\end{figure}

\autoref{fig:symm} illustrates the computational scaling with bond dimension of the \emph{ab initio} DMRG algorithm implemented in \textsc{block2} using different spin symmetries. We note that the actual scaling of \emph{ab initio} DMRG is lower than the formal \(\mathcal{O}(M^3) \) scaling (where \( M \) is the MPS bond dimension), mainly due to the use of symmetries and block-sparse matrix operations.\cite{xiang2023distributed}

\subsubsection{Sweep algorithms}

Standard DMRG optimizations (sweeps) update the wavefunction in one of two ways, termed the two-site and one-site algorithms.\cite{white1993density,white2005density,hubig2015strictly} 
The local Hilbert space in DMRG refers to the Fock space of the site, which is of, e.g.~dimension \( d=4 \) for a spatial orbital and \( d=2 \) for a spin orbital. 
Assuming each DMRG site is a spatial orbital, the two-site algorithm uses the product space of the larger two-site local Hilbert space \( d = 16 \) and the left and right block renormalized states to represent the DMRG renormalized wavefunction (which helps escape from local minima) while the one-site algorithm 
substitutes the \( d = 4 \) local Hilbert space, lowering memory and computational costs. 
An effective ``half-site'' calculation  with \( d = 2\) can also be performed by using the one-site algorithm with spin orbital sites. One can also fuse  sites in the sweep [by contracting adjacent tensors in the Hamiltonian matrix product operator (MPO)] before performing DMRG,  allowing \textsc{block2} to perform larger site sweep optimizations.

We note that the dimension of the local Hilbert space on different sites need not be the same. For example, DMRG algorithms with one or two large sites\cite{larsson2022matrix} use a single site to represent all the virtual and/or core orbitals to  efficiently treat dynamic correlation. In such cases, \textsc{block2}  supports mixing the one-site and two-site sweep algorithms to balance efficiency and accuracy.\cite{larsson2022matrix}

\subsubsection{Excited states}

The ground-state DMRG algorithm can be extended to optimize excited states and such excited-state algorithms are available in \textsc{block2}. The excited-state algorithms implemented in \textsc{block2} take the following forms:

(i) The state-averaged DMRG approach.\cite{dorando2007targeted} In 
this formalism, there is a common set of rotation matrices for all states, but the center site carries a unique tensor (renormalized wavefunction) for each state. This algorithm is normally a robust and convenient choice for tens of roots, but for a given bond dimension, the accuracy decreases as the number of roots is increased.

(ii) The projected orthogonalization approach.\cite{stackblock,keller2015efficient,larsson2019computing} To improve the quality of the states obtained from the state-averaged DMRG, one can further refine each excited state (as an independent MPS) by projecting out lower-lying states when optimizing the central site. 

(iii) The level-shift approach.\cite{wouters2014chemps2,fishman2022itensor} Here, we compute the ground and excited states one-by-one with a modified level-shifted Hamiltonian defined as
\begin{equation}
    \hat{H}'_k = \hat{H} + \sum_{i=1}^{k-1} w_i |\Psi_i\rangle\langle \Psi_i|
\end{equation}
where \( |\Psi_i\rangle \) are the converged states below the targeted excited state \( |\Psi_k\rangle \), and \( w_i \) are the energy level shifts. This can be combined with the state-averaged ansatz to determine  batches of states at a time. 

(iv) The Harmonic Davidson approach.\cite{dorando2007targeted} By changing the standard Davidson solver to the Harmonic Davidson solver, one can directly target interior excited states close to a given energy without computing all the states below. In the current implementation of \textsc{block2}, this approach is not as stable as the above approaches for dense spectra.

\subsubsection{Relativistic effects}

Relativistic effects can be classified into scalar relativistic and spin-dependent relativistic effects,\cite{li2014spin} where
the mean-field level scalar relativistic effects can be described inexpensively.\cite{moritz2005relativistic}
\textsc{Block2} supports several strategies to include the scalar and spin-dependent relativistic effects.

(i) Relativistic DMRG.\cite{knecht2014communication,battaglia2018efficient,hoyer2022relativistic} We can directly solve for eigenstates of four-component Dirac-Coulomb, Dirac-Coulomb-Gaunt, or Dirac-Coulomb-Breit Hamiltonians.\cite{dyall2007introduction} Here DMRG is performed in complex number mode where both the  MPS and MPO are complex. 

(ii) One-step SOC-DMRG.\cite{zhai2022comparison} Within  the two-component formalism of relativistic quantum chemistry, we can include spin-orbit coupling (SOC) by adding a SOC term to the scalar-relativistic Hamiltonian, and then perform DMRG on this two-component Hamiltonian with a complex MPO and MPS. Alternatively, we can simplify the SOC term within the spin-orbit mean-field (SOMF) approximation,\cite{neese2005efficient} giving rise to complex one-electron and real two-electron integrals. One can then represent the Hamiltonian as the sum of two MPOs where the expensive two-electron interaction is contained only in the second MPO, which is completely real.\cite{zhai2022comparison} One then runs \textsc{block2} in the hybrid complex and real arithmetic DMRG mode, where the real and imaginary parts of the effective wavefunction share the same real rotation matrices in the MPS.

(iii) Two-step SOC-DMRG.\cite{roemelt2015spin,sayfutyarova2016state,sayfutyarova2018electron} When the SOC effect is weak, we can  first compute eigenstates of the scalar-relativistic Hamiltonian. Then SOC effects can be added by carrying out the state-interaction of a small number of eigenstates.

\begin{figure}[!htbp]
  \includegraphics[width=\columnwidth]{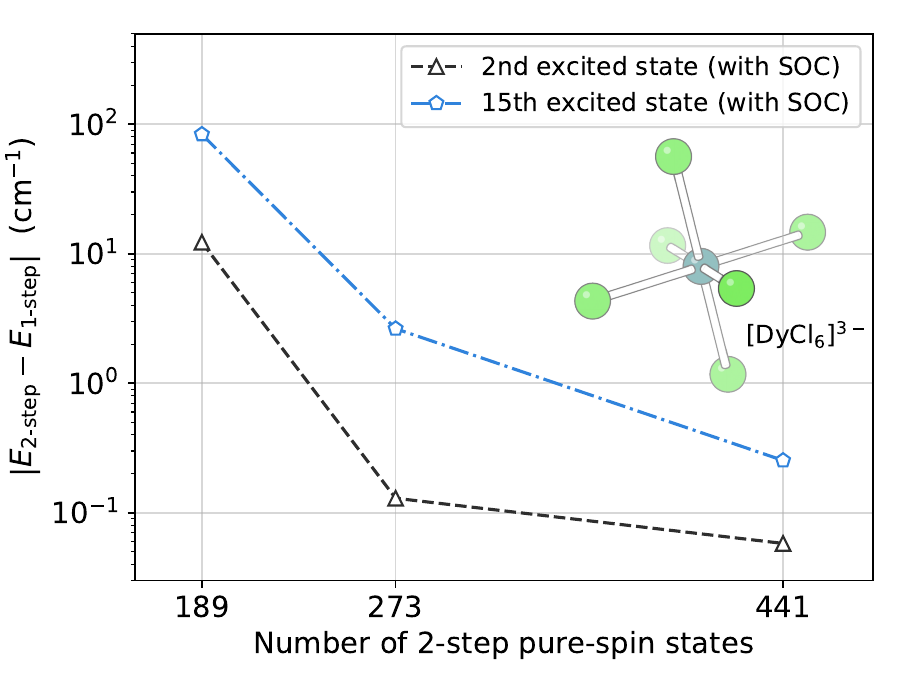}
  \caption{The energy difference between representative excited states computed from the 1-step and 2-step SOC-DMRG approaches, for the (7o, 9e) active space of the  \( [\mathrm{DyCl_6}]^{3-} \) cluster. Adapted from Fig. 4(a) in \lit{zhai2022comparison}.}
  \label{fig:soc}
\end{figure}

\autoref{fig:soc} shows a numerical comparison of the  accuracy of the 1-step and 2-step SOC-DMRG treatments.

\subsubsection{Non-standard Hamiltonians and operators}

In addition to computations involving the standard Hermitian quantum chemistry Hamiltonian with up to two-body interaction terms, \textsc{block2} supports the use of other types of non-standard Hamiltonians and operators in DMRG algorithms and/or MPO/MPS operations.

(i) General Hamiltonians with high-order interactions. Using the techniques in Section~\ref{sec:mpoconstruction}, \textsc{block2} supports DMRG computations with arbitrary parity-preserving (i.e. even number of creation and annihilation operator) Hamiltonians. For example, 
DMRG calculations can be carried out for Hamiltonians with three-body or up to $n$-body terms. We note that higher than two-body terms appear commonly in quantum chemistry in effective Hamiltonians obtained from similarity transformations\cite{chan2005density} or canonical transformations \cite{white2002numerical,yanai2006canonical,neuscamman2010strongly,yanai2012extended} of the \emph{ab initio} Hamiltonians.

(ii) Normal-ordered Hamiltonians.\cite{bartlett2007coupled} The DMRG algorithm can work instead with the normal-ordered version of a Hamiltonian where the vacuum expectation value is separated out. This can make the DMRG algorithm with reduced floating point precision numerically more stable.

(iii) Non-Hermitian Hamiltonians.\cite{chan2005density,mitrushenkov2003possibility,liao2023density,baiardi2020transcorrelated} This allows DMRG to be used, for example, with Hamiltonians that arise from many-body similarity transformations, such as in coupled cluster theory or the transcorrelated method.

(iv) Anti-Hermitian operators. \textsc{block2} has special support for anti-Hermitian operators. Exponentials of anti-Hermitian operators appear in various applications, most notably, when performing orbital rotations of the MPS, or in canonical transformation theory.\cite{yanai2006canonical} The explicit support for anti-Hermitian operators allows such unitaries to be expressed as an exponential of a real MPO, thus avoiding complex arithmetic.

(v) General MPO and MPS operations with arbitrary operators. \textsc{block2} supports MPO and MPS operations with arbitrary operators, including non-number-conserving operators, as described in Section~\ref{sec:mpomps}. For example, to support the MPS compression approach for strongly-correlated NEVTP2,\cite{sokolov2017time} one needs to apply a partial Hamiltonian to the reference wavefunction to get the perturber wavefunction
\begin{equation}
    |\Psi'_i\rangle = \bigg( \sum_{abc} v_{iabc} \ a^\dagger_a a^\dagger_b a_c \bigg) \ |\Psi_0\rangle
\end{equation}
In \textsc{block2} this is done by building an MPO for the operator part (as indicated by the parenthesis) in the above expression.

\subsubsection{Initial guesses}

\textsc{block2} provides several ways to generate initial MPS guesses for various algorithms. It also provides an implementation of perturbative noise\cite{white2005density,hubig2015strictly} which helps DMRG algorithms escape from poor initial guesses. 

(i) A random MPS with the desired bond dimension can be generated.\cite{wouters2014density}

(ii) One can supply an estimate of the occupancy information at each site from a cheaper method [e.g.~coupled cluster singles and doubles (CCSD)]. \textsc{block2} then uses this information to construct the initial quantum number distribution in the MPS (by assuming that the probability distribution of the occupancy at each site is independent  from each other).

(iii) We can generate an initial guess from another (more approximate) MPS using fitting. For example, we can use an MPS parameterized and optimized in the singles and doubles excitation space to initialize the MPS in the full Hilbert space.

(iv) We can construct an initial guess MPS from a given linear combination of determinants or Configuration State Functions (CSFs).

\subsubsection{Orbital reordering algorithms}

For a given bond dimension, the accuracy of DMRG is affected by the orbital ordering.\cite{moritz2005convergence} \textsc{block2} provides several options to choose good orbital orderings in DMRG.

(i) For molecules with non-trivial point group symmetries, it can group the orbitals by irreducible representation.\cite{ma2013assessment,mishmash2023hierarchical}

(ii) For less symmetric systems, it can choose an ordering using the Fiedler vector\cite{fiedler1975property} of an appropriate metric matrix. This ordering method is very inexpensive.\cite{barcza2011quantum,olivares2015ab}

(iii) The Fiedler ordering can be further optimized using a more expensive genetic algorithm (GA),\cite{olivares2015ab} as used in the \textsc{StackBlock} code.\cite{stackblock} In \textsc{block2}, this GA is reimplemented with greatly increased efficiency.

The Fiedler and GA orderings require one to define a metric to measure the correlation between pairs of orbitals. \textsc{block2} can use the absolute value of the exchange matrix\cite{olivares2015ab} or the two-orbital mutual information [computed from the one- and two-orbital density matrices (1 and 2ODMs)]\cite{rissler2006measuring} for this purpose.

\subsubsection{DMRG-CASSCF}

For typical quantum chemistry problems, it is often not possible to treat all orbitals in DMRG. One can combine DMRG with the Complete Active Space Self-Consistent Field (CASSCF) method\cite{sun2017general,smith2022near} to use DMRG to approximate the FCI problem within the active space while optimizing the active space orbitals.\cite{zgid2008density,ghosh2008orbital,yanai2009accelerating,hu2015excited,iino2023algorithm} Using \textsc{block2} we can perform DMRG-CASSCF through interfaces provided by external quantum chemistry packages, such as \textsc{PySCF}.\cite{sun2018pyscf,sun2020recent} With \textsc{PySCF} we currently support DMRG-CASSCF energy optimization\cite{ghosh2008orbital} for spin restricted and spin unrestricted orbitals, and DMRG-CASSCF gradients and geometry optimization\cite{hu2015excited} for spin restricted orbitals.

\subsection{High performance DMRG}

Parallelization in \emph{ab initio} DMRG is essential to realistic chemical problems. In \textsc{block2} we have implemented a hierarchy of different parallel and related DMRG strategies. 

\subsubsection{Standard parallelization strategies}

In a recent report,\cite{zhai2021low} we summarized five standard parallel DMRG strategies implemented in \textsc{block2}: (i) parallelism over the sum of sub-Hamiltonians,\cite{chan2016matrix} (ii) parallelism over sites,\cite{stoudenmire2013real,chen2021real} (iii) parallelism over normal and complementary operators,\cite{chan2004algorithm,wouters2014density} (iv) parallelism over symmetry sectors,\cite{kurashige2009high,levy2020distributed} and (v) parallelism within dense matrix multiplications.\cite{hager2004parallelization,levy2020distributed} For large scale calculations, it is essential to combine these strategies to achieve high performance.\cite{zhai2021low}

We have implemented most of these strategies in \textsc{block2} not only for the ground state optimization problem, but also for other types of calculations, such as time evolution and the calculation of response properties. The user can change the parallelization strategy at runtime to maximize the efficiency of custom workflows

\begin{figure}[!htbp]
  \includegraphics[width=\columnwidth]{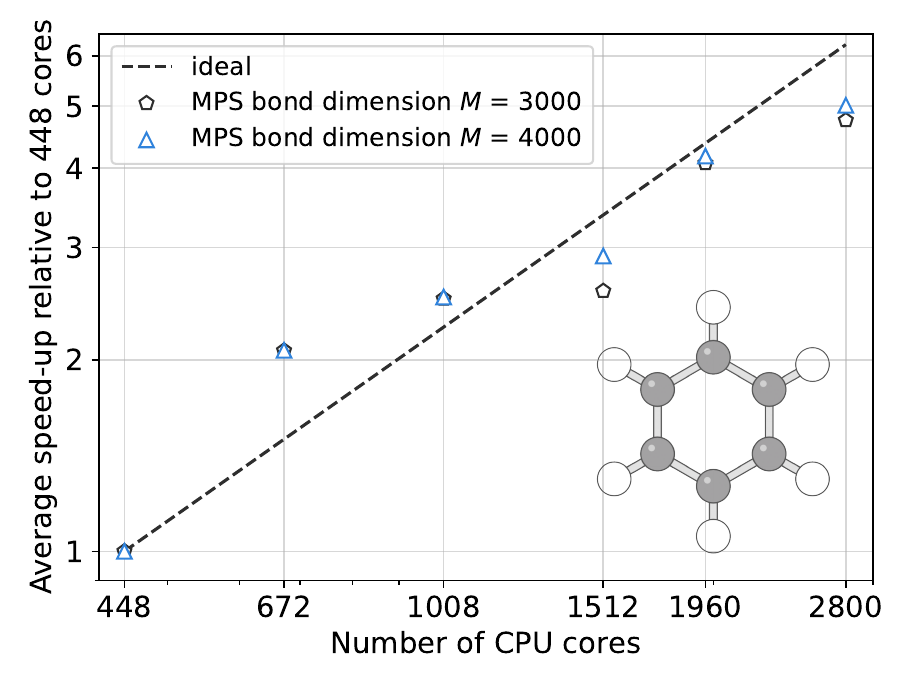}
  \caption{Speed-up of average wall time per sweep relative to the \( N_{\mathrm{core}} = 448 \) case for different MPS bond dimensions using parallel DMRG algorithms, for solving the ground state in a (108o, 30e) active space of the benzene molecule\cite{eriksen2020ground} using a cc-pVDZ basis.\cite{dunning1989gaussian} Adapted from Fig. 4 in \lit{zhai2021low}.}
  \label{fig:parallel}
\end{figure}

\autoref{fig:parallel} shows a numerical example of the performance of parallel DMRG.

\subsubsection{Integral distribution strategies}

In large \emph{ab initio} calculations, the memory to store the two-electron integrals of the Hamiltonian can be a bottleneck. In \textsc{block2} we provide several integral/MPO distribution strategies to balance the CPU, disk, and memory consumption in such calculations.

(i) In the default strategy for small- to medium-sized problems, integrals are replicated across different nodes, which provides the most flexibility for the DMRG algorithm. In this case, we have the freedom to switch between the normal/complementary (NC) parallelization strategy and the complementary/normal (CN) parallelization strategy near the middle site of a sweep,\cite{kurashige2009high} which leads to higher computational efficiency.

(ii) For larger calculations where integral replication is problematic, we provide an alternative strategy where each node only stores the portion of integrals required by that node. This strategy reduces the communication and memory costs, at the price of slightly reduced CPU efficiency.\cite{zhai2021low}

(iii) When the number of orbitals is large and localization has been performed, there are many  small or close-to-zero integral elements. \textsc{block2} can store the integrals in a compressed format to further reduce memory consumption, using a floating point number compression algorithm introduced in \lit{zhai2021low}. (Note that this only affects the memory consumption before the Hamiltonian MPO is built.) 

(iv) For even larger scale applications, the MPOs constructed from the sub-Hamiltonians can also consume a large amount of memory on each node (since each MPO must contain at least the same amount of information as contained in the integrals). To alleviate this problem, we can store the MPO primarily on disk and only the MPO tensors required at the given one or two sites in an iteration of the sweep algorithm are loaded into memory. 
Note that MPO storage is not a problem in traditional DMRG implementations like \textsc{StackBlock} where the action of the MPO is performed on the fly.

(v) At the cost of introducing some error, the bond dimension of the sub-Hamiltonian MPOs can be compressed during the construction as described in Section~\ref{sec:mpoconstruction}. 

\subsubsection{Parallelization strategies for specific tasks}

The aforementioned parallelization strategies are common to a wide range of DMRG algorithms. We have additionally implemented more specialized parallelization strategies for some specific computational tasks.

(i) The generation of the Hamiltonian matrix elements in large-site DMRG calculations (used e.g.~to treat multireference dynamical correlation\cite{larsson2022matrix}) is implemented with shared-memory parallelization over the determinants or CSFs.

(ii) Green's function calculations in dynamical DMRG\cite{ronca2017time} are parallelized in a distributed manner over the site index of the one-particle Green's function matrix element.

(iii) Stochastic perturbative DMRG\cite{guo2018communication} is implemented with a hybrid shared-memory and distributed parallelization to sample determinants or CSFs.

(iv) The GA algorithm for orbital reordering \cite{olivares2015ab} is implemented with distributed parallelization over the different initial populations in a multi-start style GA algorithm.

(v) The MPS compression step in the DMRG version of second order strongly-contracted \(N\)-Electron Valence States for Multireference Perturbation Theory (SC-NEVPT2)\cite{sokolov2017time} is implemented with distributed parallelization over virtual orbital indices.

\subsection{General operators and MPO construction algorithms}

\label{sec:mpoconstruction}

\textsc{block2} provides full support for algorithms involving general Hamiltonians (and other operators) expressed as MPOs. The implementation in \textsc{block2} is carefully designed to support such general operations while not (severely) sacrificing efficiency. 
As discussed in \lit{chan2016matrix}, the original \emph{ab initio} DMRG algorithm\cite{white1999ab} implicitly provides an efficient implementation of the expectation value of a sparse MPO, but does not explicitly construct the MPO. The advantages of an explicit MPO object for relevant operators and Hamiltonians is that  a wider variety of algorithms can be easily implemented. However, naive implementations of the MPO for complicated operators such as the \emph{ab initio} Hamiltonian lead to the wrong cost scaling.\cite{chan2016matrix} Consequently, \textsc{block2} uses the techniques described below to achieve high efficiency.

\subsubsection{Symbolic MPO approach for \emph{ab initio} Hamiltonians}

To obtain performance competitive with traditional implementations of \emph{ab initio} DMRG (such as the one found in \textsc{StackBlock}),\cite{chan2002highly,sharma2012spin} \textsc{Block2} provides a symbolic MPO class for \emph{ab initio} Hamiltonians. Here, the multiplication of the MPO matrices is interpreted as a set of formulae that reproduce the DMRG equations in traditional \emph{ab initio} implementations. Symbolic MPOs are available for different spin symmetries: $S^2$ (SU2), $S_z$, and no spin symmetry, and support the Normal/Complementary (NC) partition, the Complementary/Normal (CN) partition, and the more efficient mixed partition with a NC to CN transformation near the middle site.\cite{kurashige2009high} In addition,
simplification rules are applied during the symbolic multiplication of the MPO matrices
to use the index permutation and transposition symmetry of the normal and complementary operators\cite{kurashige2009high} (such as \( \hat{A}_{ij} = -\hat{A}_{ji} \) where \( \hat{A}_{ij} \) is defined as \( a_i^\dagger a_j^\dagger \)). This reduces the number of non-redundant renormalized operators that have to be stored and computed during DMRG sweeps. With this optimization, 
the DMRG algorithm implemented via the symbolic MPO is more memory efficient than with MPOs constructed using other approaches.
In addition, because the symbols correspond to the labels of renormalized operators in a traditional DMRG algorithm, we can also implement conventional DMRG parallelization strategies in the literature within this MPO approach.\cite{chan2004algorithm,chan2016matrix}

\subsubsection{General MPO construction for Hamiltonians}

\textsc{block2} also implements general operator and Hamiltonian MPOs efficiently.
At the interface level, the user writes the definition of the operator as a symbolic expression, using characters in a string to represent elementary operators such as \( \hat{a}^\dagger \) and \( \hat{a} \). If non-standard elementary operators are required, the user can define the matrix representation of each elementary operator in a Python script. 
Bosonic operators\cite{jeckelmann1998density,ge2022computational} and general Pauli strings\cite{mishmash2023hierarchical} are supported. Starting from such an arbitrary symbolic expression of a many-body operator (such as the Hamiltonian), and the matrix representation of the involved elementary operators (in any local orthonormal basis), the code generates the MPO automatically. One can then use the MPO to optimize or evaluate expectation values, and carry out other DMRG computations. Both spin-adapted and non-spin-adapted MPOs can be generated. \autoref{lst:script} is an example Python script for performing DMRG of the Hubbard-Holstein model using \textsc{block2}.

\onecolumngrid
\begin{lstlisting}[label={lst:script},caption={A Python script to setup the Hubbard-Holstein model Hamiltonian and perform DMRG to find the ground state using \textsc{block2}. In the script, we use the characters \texttt{c, d, C, D, E,} and \texttt{F} (as an example) to represent the fermionic elementary operators \( a_\alpha^\dagger, a_\alpha, a_\beta^\dagger,\) and \( a_\beta\) and the bosonic elementary operators \( b^\dagger \) and \( b \), respectively. The expected output of this script (the ground state energy) is \(-6.956893\).}]
from pyblock2.driver.core import DMRGDriver, SymmetryTypes, MPOAlgorithmTypes
import numpy as np

N_SITES_ELEC, N_SITES_PH, N_ELEC = 4, 4, 4
N_PH, U, OMEGA, G = 11, 2, 0.25, 0.5
L = N_SITES_ELEC + N_SITES_PH

driver = DMRGDriver(scratch="./tmp", symm_type=SymmetryTypes.SZ, n_threads=4)
driver.initialize_system(n_sites=L, n_elec=N_ELEC, spin=0)

# [Part A] Set states and matrix representation of operators in local Hilbert space
site_basis, site_ops = [], []
Q = driver.bw.SX # quantum number wrapper (n_elec, 2 * spin, point group irrep)

for k in range(L):
    if k < N_SITES_ELEC:
        # electron part
        basis = [(Q(0, 0, 0), 1), (Q(1, 1, 0), 1), (Q(1, -1, 0), 1), (Q(2, 0, 0), 1)] # [0ab2]
        ops = {
            "": np.array([[1, 0, 0, 0], [0, 1, 0, 0], [0, 0, 1, 0], [0, 0, 0, 1]]),   # identity
            "c": np.array([[0, 0, 0, 0], [1, 0, 0, 0], [0, 0, 0, 0], [0, 0, 1, 0]]),  # alpha+
            "d": np.array([[0, 1, 0, 0], [0, 0, 0, 0], [0, 0, 0, 1], [0, 0, 0, 0]]),  # alpha
            "C": np.array([[0, 0, 0, 0], [0, 0, 0, 0], [1, 0, 0, 0], [0, -1, 0, 0]]), # beta+
            "D": np.array([[0, 0, 1, 0], [0, 0, 0, -1], [0, 0, 0, 0], [0, 0, 0, 0]]), # beta
        }
    else:
        # phonon part
        basis = [(Q(0, 0, 0), N_PH)]
        ops = {
            "": np.identity(N_PH), # identity
            "E": np.diag(np.sqrt(np.arange(1, N_PH)), k=-1), # ph+
            "F": np.diag(np.sqrt(np.arange(1, N_PH)), k=1),  # ph
        }
    site_basis.append(basis)
    site_ops.append(ops)

# [Part B] Set Hamiltonian terms in Hubbard-Holstein model
driver.ghamil = driver.get_custom_hamiltonian(site_basis, site_ops)
b = driver.expr_builder()

# electron part
b.add_term("cd", np.array([[i, i + 1, i + 1, i] for i in range(N_SITES_ELEC - 1)]).ravel(), -1)
b.add_term("CD", np.array([[i, i + 1, i + 1, i] for i in range(N_SITES_ELEC - 1)]).ravel(), -1)
b.add_term("cdCD", np.array([[i, i, i, i] for i in range(N_SITES_ELEC)]).ravel(), U)

# phonon part
b.add_term("EF", np.array([[i + N_SITES_ELEC, ] * 2 for i in range(N_SITES_PH)]).ravel(), OMEGA)

# interaction part
b.add_term("cdE", np.array([[i, i, i + N_SITES_ELEC] for i in range(N_SITES_ELEC)]).ravel(), G)
b.add_term("cdF", np.array([[i, i, i + N_SITES_ELEC] for i in range(N_SITES_ELEC)]).ravel(), G)
b.add_term("CDE", np.array([[i, i, i + N_SITES_ELEC] for i in range(N_SITES_ELEC)]).ravel(), G)
b.add_term("CDF", np.array([[i, i, i + N_SITES_ELEC] for i in range(N_SITES_ELEC)]).ravel(), G)

# [Part C] Perform DMRG
mpo = driver.get_mpo(b.finalize(adjust_order=True), algo_type=MPOAlgorithmTypes.FastBipartite)
mps = driver.get_random_mps(tag="KET", bond_dim=250, nroots=1)
energy = driver.dmrg(mpo, mps, n_sweeps=10, bond_dims=[250] * 4 + [500] * 4,
    noises=[1e-4] * 4 + [1e-5] * 4 + [0], thrds=[1e-10] * 8, dav_max_iter=30, iprint=2)
print("DMRG energy = %20.15f" % energy)

\end{lstlisting}
\twocolumngrid

The key to achieving efficiency is to construct MPOs in such a way as to obtain sparse MPO tensors. We note that MPOs contain a gauge degree of freedom and can thus be written in multiple ways, but a sparse representation 
 is typically available as the operators of interest are quite restricted (e.g.~of one- and two-particle form). To correctly identify a sparse MPO representation, several ideas have been discussed in the literature.\cite{keller2015efficient,chan2016matrix,hubig2017generic,ren2020general} In \textsc{block2}, we implement two of them.

(i) The bipartite approach.\cite{ren2020general} This is an efficient graph-theory-based approach to build an exact MPO that identifies the best sparse representation. When used with the \emph{ab initio} Hamiltonian, this produces the optimal MPO bond dimension and sparse structure of the matrices, and thus the correct complexity in an \emph{ab initio} DMRG algorithm.

\begin{figure}[!htbp]
  \includegraphics[width=\columnwidth]{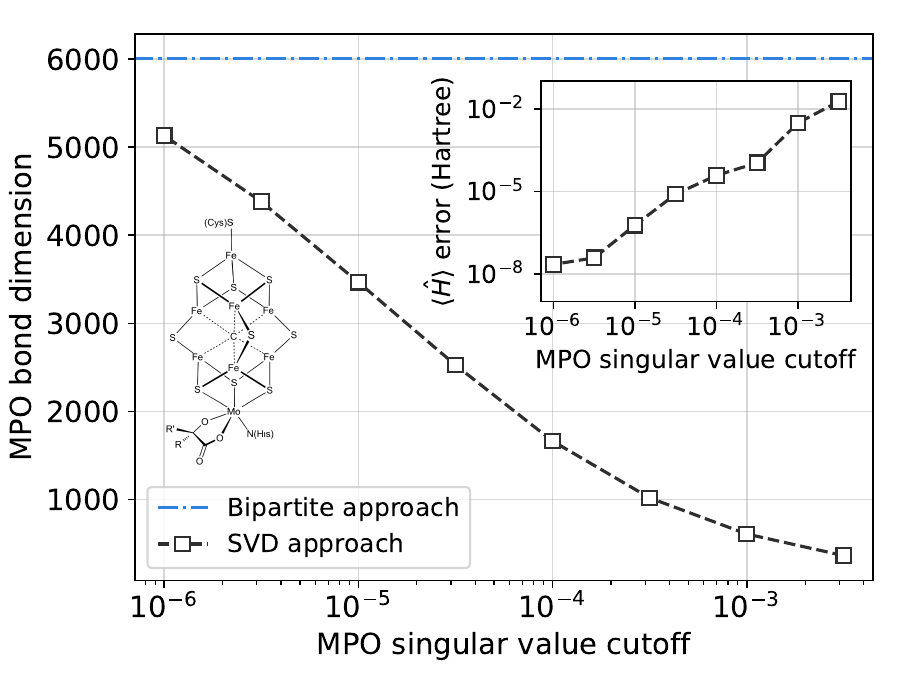}
  \caption{A comparison between the MPO bond dimensions for MPOs constructed using the bipartite approach\cite{ren2020general} and the SVD approach\cite{chan2016matrix} with different cutoffs for the singular values, for the (76o, 113e) active space model of the FeMo cofactor.\cite{kirn1992crystallographic} The active space model is from \lit{li2019electronic}. The inset plot shows the error in the expectation value \( \langle \mathrm{MPS} | \mathrm{MPO} | \mathrm{MPS} \rangle \) computed using MPOs constructed with different cutoffs and an MPS (optimized using the exact MPO) with bond dimension \( M=2000 \) for the FeMo cofactor system. In the SVD approach, the cutoff was applied to the unrescaled singular values of the unnormalized Hamiltonian integrals.}
  \label{fig:mpo}
\end{figure}

(ii) The singular value decomposition (SVD) approach.\cite{chan2016matrix,stoudenmire2017sliced,lin2021low} 
This approach introduces the possibility of MPO bond dimension compression. For some operators, e.g.~\emph{ab initio} Hamiltonians of large molecules, the MPO bond dimension reduction can greatly increase the DMRG efficiency (see \autoref{fig:mpo}). For large numbers of orbitals and/or high-order (more than three-body) terms in the Hamiltonian, the direct SVD of integral matrices can be expensive, and we alleviate this problem by performing SVD for each symmetry block of the MPO tensors separately. Although the SVD does not preserve sparsity, sparsity can be introduced using the sub-Hamiltonian  strategy,\cite{chan2016matrix} which recovers the correct DMRG complexity and MPO tensor sparsity.

\autoref{fig:mpo} shows a numerical example comparing MPOs resulting from different constructions. 

\subsection{MPO/MPS algebra}

\label{sec:mpomps}

The MPO and MPS representation of operators and states allow one to view DMRG algorithms in a very general sense as an extension
of linear algebra.\cite{chan2016matrix} \textsc{block2} supports this general perspective with a convenient interface to MPO/MPS algebraic operations and algorithms. The interface dispatches at the backend to several implementations for different levels of efficiency.

\subsubsection{Sweep algorithms for MPO/MPS algebra}

Consider a general MPO/MPS algebraic expression
\begin{equation}
    |\mathrm{MPS}'\rangle =
    \operatorname{poly}(\mathrm{MPO}) \big( \alpha |\mathrm{MPS_1}\rangle + \beta |\mathrm{MPS_2}\rangle \big)
\end{equation}
where \( \operatorname{poly}(\mathrm{MPO})\) (can be a linear combination of powers of an MPO or its inverse),  \( |\mathrm{MPS_1}\rangle, \) and \( |\mathrm{MPS_2}\rangle \) are known quantities, and \( \alpha \) and \( \beta \) are known scalar numbers. In one MPO/MPS algebra backend, we numerically find \( |\mathrm{MPS}'\rangle \) by fitting. Namely, starting from a trial \( |\mathrm{MPS}'\rangle \), we solve the optimization problem
\begin{equation}
    \max_{|\mathrm{MPS}'\rangle}
    \lVert\langle \mathrm{MPS}' | \operatorname{poly}(\mathrm{MPO}) \big( \alpha |\mathrm{MPS_1}\rangle + \beta |\mathrm{MPS_2}\rangle \big) \rVert
\end{equation}
by updating the tensors in \( |\mathrm{MPS}'\rangle \) using the one-site or two-site sweep algorithms. Due to its  similarity to the standard DMRG algorithm, this procedure is implemented very efficiently in \textsc{block2}. 
The standard caveats of DMRG optimization apply however, namely, there can be a dependence on the initial guess (especially for small bond dimension) and the global optimality of the solution is not guaranteed. 
Using this approach, \textsc{block2} implements the following important primitives of MPO/MPS algebra.

(i) Addition, subtraction, and scalar multiplication of MPS. This is implemented by fitting \( |\mathrm{MPS}'\rangle \) to \( \alpha |\mathrm{MPS_1}\rangle + \beta |\mathrm{MPS_2}\rangle \).

(ii) Matrix-vector multiplication. This is implemented by fitting \( |\mathrm{MPS}'\rangle \) to \( \mathrm{MPO} \ |\mathrm{MPS}\rangle \).

(iii) Linear equation or matrix inversion. This can be formally expressed as \( |\mathrm{MPS}'\rangle = \mathrm{MPO}^{-1} |\mathrm{MPS}\rangle\). By using sweep algorithms,\cite{jeckelmann2002dynamical,sharma2014communication,ronca2017time} we can transform this equation into an effective linear algebra equation defined on one or two sites of \( |\mathrm{MPS}'\rangle \), which can be solved using iterative algorithms such as Conjugate Gradient (CG),\cite{shewchuk1994introduction} MINRES,\cite{harman1966factor}, GCROT(\(m,k\)),\cite{hicken2010simplified}, IDRS,\cite{van2011algorithm} Chebyshev,\cite{mohr2017efficient} and LSQR.\cite{paige1982lsqr}

(iv) Matrix exponentiation. This can be formally expressed as \( |\mathrm{MPS}'\rangle = \exp (-\beta \ \mathrm{MPO})\ |\mathrm{MPS}\rangle\),
 and \( |\mathrm{MPS}'\rangle \) can be found using the time-dependent DMRG methods described in Section~\ref{sec:tddmrg}.

\subsubsection{Exact algorithms for MPO/MPS algebra}

MPO/MPS algebra can also be implemented exactly using block-sparse tensor contraction operations. For typical quantum chemistry problems the bond dimensions of the MPS and MPOs are on the order of several thousand, and intermediates generated by the contraction of MPS and MPO tensors can quickly exceed the available memory. Nevertheless, this scheme can be useful for model Hamiltonians with simpler interactions or for benchmarking purposes.

To support exact MPO/MPS algebra, we provide a separate open source library called \textsc{pyblock3}\cite{pyblock3} which more fully exposes the tensor network view of DMRG algorithms, at some cost to efficiency.
In \textsc{pyblock3}, the MPO, MPS, and block-sparse tensor objects can be easily manipulated with an interface similar to performing operations on \textsc{numpy} arrays,\cite{harris2020array} so that arbitrary MPO/MPS algebra 
is straightforward. We provide subroutines to translate MPO and MPS objects back and forth between \textsc{block2} and \textsc{pyblock3}. Note that we do not currently support SU(2) symmetry in \textsc{pyblock3}, thus this interoperability is only possible for non-spin-adapted objects.

As MPS and MPO are represented as pure Python data structures in \textsc{pyblock3}, users can further explore many other possibilities through the interface between \textsc{block2} and \textsc{pyblock3}. For example, utilizing the automatic differentiation\cite{paszke2017automatic} feature implemented in \textsc{pyblock3}, we can perform gradient style optimization of MPO or MPS tensors.

\subsubsection{Symbolic manipulation and simplification of operator expressions}

For algebra involving only fermionic operators, 
\textsc{block2} provides an efficient symbolic engine to manipulate, simplify, and truncate expressions  using Wick's theorem.\cite{bartlett2007coupled} Compared to other existing similar implementations,\cite{hirata2003tensor,hirata2006symbolic,neuscamman2009quadratic,saitow2013multireference,macleod2015communication,saitow2015fully,evangelista2022automatic} we provide a user friendly Python interface with support for expressions arising in spin-free, spin unrestricted, and general spin theories. Once the algebraic simplifications have been performed, the resulting symbolic expression can be converted into an MPO using techniques in Section~\ref{sec:mpoconstruction}. Some examples of applications include:

(i)  MPO construction of the normal-ordered \( \hat{H} \). Such MPOs increase the numerical stability in low precision DMRG.

(ii) MPO construction of \( \hat{H}^2 \), used to compute the energy variance of an MPS. Note that computing the norm of \(\hat{H}|\mathrm{MPS}\rangle\) (by first constructing an MPS intermediate of \(\hat{H}|\mathrm{MPS}\rangle\) using fitting) is an alternative way to obtain the variance which is also
 often cheaper. 

(iii) MPO construction of \( \mathrm{e}^{-\hat{T}}\hat{H}\mathrm{e}^{\hat{T}} \), which arises in theories involving a similarity transform of the Hamiltonian.\cite{bartlett2007coupled} In practice, we first rewrite the expression in terms of nested commutators using the Baker-Campbell-Hausdorff expansion. After simplification (and truncation if necessary) we obtain an expression that can be used to generate the MPO.

\begin{figure}[!htbp]
  \includegraphics[width=\columnwidth]{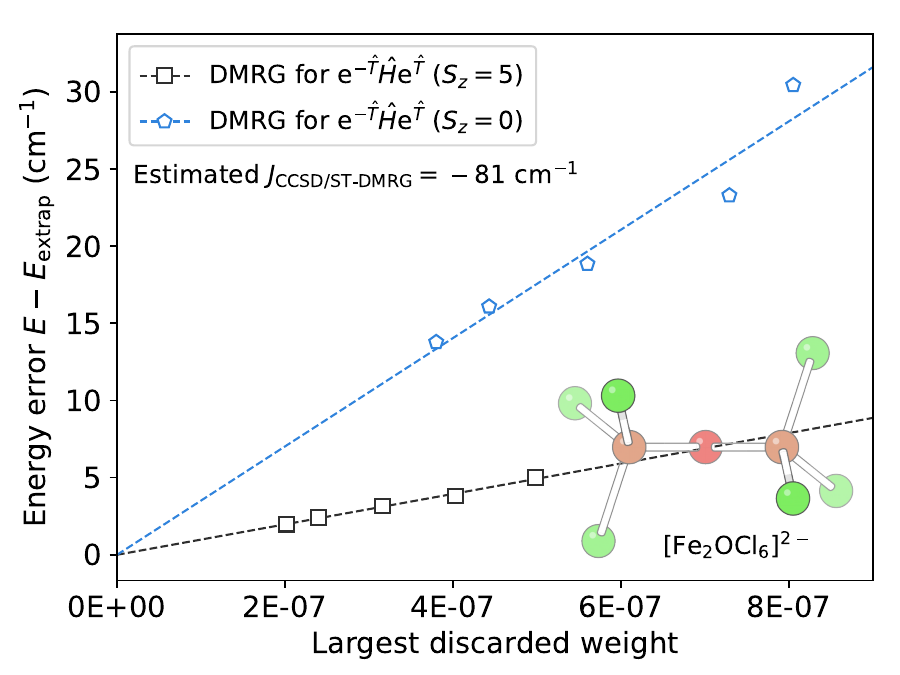}
  \caption{The extrapolation of DMRG sweep energies using MPS bond dimension \(M = 600, 500, 400, 300,\) and \( 200 \), for the (30o, 30e) active space of the similarity transformed Hamiltonian \( \mathrm{e}^{-\hat{T}}\hat{H}\mathrm{e}^{\hat{T}} \) with only terms up to three-body interaction kept after normal ordering, computed for the high spin (\(S_z = 5\)) and low spin (\(S_z = 0\)) states of the \( [\mathrm{Fe_2OCl_6}]^{2-} \) cluster, using the cc-pVDZ-DK basis and the exact two-component relativistic correction.\cite{saue2011relativistic} The cluster geometry is from \lit{schurkus2020theoretical}.}
  \label{fig:st}
\end{figure}

\autoref{fig:st} shows a numerical example of  DMRG using a similarity transformed Hamiltonian constructed from CC amplitudes.

\subsection{MPS analysis and properties}

Once we have computed an MPS (e.g.~from the DMRG sweep algorithm) we can analyze it in terms of quantities familiar to standard quantum chemistry. Below, we list some representative quantities that can be extracted from an MPS (or a pair of bra and ket MPSs) in \textsc{block2}.

\subsubsection{$N$-particle reduced density matrices}

\(N\)-particle density matrices (\(N\)-PDMs)\cite{zgid2008obtaining,ghosh2008orbital,guo2016n} summarize the information in a wavefunction and have many applications across quantum chemistry, such as in  orbital optimization,\cite{zgid2008density,ghosh2008orbital} (internally) contracted dynamical correlation theories,\cite{saitow2015fully, guo2016n} and so on. In \textsc{block2} we provide an efficient algorithm to compute \( N\)-PDMs of MPS with arbitrary \( N \) (in practice, only \( N < 5 \) can be computed in non-trivial systems). The implementation has the following features.

(i) We support both \( N\)-PDMs (with integer \( N \)) and transition \( N\)-PDMs (with integer or half-integer \( N \)). In a unified framework, the user can compute spin-traced, spin specific, or general spin PDMs.

(ii) The user can freely choose the fermionic operator ordering in the \( N\)-PDMs via an operator string input parameter. For example, for the spin-traced 3-PDM, both
\begin{equation}
    E_{abcdef}^{(1)} := \sum_{\sigma\tau\lambda} \langle
    a^\dagger_{a\sigma} a^\dagger_{b\tau} a^\dagger_{c\lambda} a_{d\lambda} a_{e\tau} a_{f\sigma} \rangle
\end{equation}
and
\begin{equation}
    E_{abcdef}^{(2)} := \sum_{\sigma\tau\lambda} \langle
    a^\dagger_{a\sigma} a_{b\sigma} a^\dagger_{c\tau} a_{d\tau} a^\dagger_{e\lambda} a_{f\lambda} \rangle
\end{equation}
notations are supported, where the former is compatible with the \textsc{StackBlock} convention,\cite{stackblock} and the latter is used in the literature of spin-free NEVPT2.\cite{angeli2002n} 

(iii) When obtaining multiple \( N\)-PDMs or PDMs with different operator orderings, they can be computed simultaneously within a single sweep so that intermediates can be reused, saving computational time.

(iv) When only a part of the PDM is required (by restricting the indices) we can compute only the required part with a greatly reduced complexity. For example, the diagonal 2-PDM
\begin{equation}
    E_{ppqq} := \sum_{\sigma\tau} \langle
    a^\dagger_{p\sigma} a_{p\sigma} a^\dagger_{q\tau} a_{q\tau}
    \rangle
\end{equation}
required in perturbative DMRG\cite{guo2018communication} can be computed with a complexity similar to that of the 1-PDM.

(v) We use a hybrid distributed and shared-memory parallelization strategy. For the shared-memory part, we use parallelism over non-redundant collective matrix element indices. For the distributed part, we use a hybrid strategy based on parallelism over sub-operator-strings and parallelism over single matrix element indices.

\subsubsection{Entanglement analysis}

We can compute various entanglement metrics of interest from the MPS.\cite{legeza2003optimizing,rissler2006measuring,barcza2011quantum,boguslawski2012entanglement,boguslawski2013orbital}

(i) One can access the bipartite entanglement entropy\cite{legeza2003optimizing,liu2012bipartite} in the MPS at each site after a standard DMRG optimization.

(ii) The 1 and 2 orbital DMs\cite{boguslawski2013orbital} (ODMs) can be computed using the general \( N \)-PDM interface implemented in \textsc{block2}, with a computational scaling similar to that of the 2-PDM. The mutual information matrix of orbital interactions can be constructed from these ODMs, as a measure of the entanglement between pairs of orbitals.\cite{boguslawski2012entanglement} The mutual information can then be used as the cost function in GA,\cite{olivares2015ab} to find an optimal orbital ordering for DMRG. Currently this feature is only implemented for non-spin-adapted MPS.

\subsubsection{Coefficients of configuration state functions or determinants}

In quantum chemistry applications, it is useful to understand the determinant or CSF expansion of the DMRG wavefunction (MPS) and these can be computed using \textsc{block2}. In the following we provide some representative use cases:

(i) In \textsc{block2}, one can extract all CSFs or determinants and their coefficients where the absolute value of the coefficient in the CI expansion is above a given threshold. This uses a depth-first search algorithm performed along the sweep.\cite{lee2021externally} When only a few CSFs or determinants satisfy the criterion, the computational cost of this algorithm is almost negligible. From the spin coupling pattern in the dominant CSFs or determinants, one obtains valuable insights into the electronic structure. Currently, for spin-adapted MPS with a non-zero total spin, the algorithm is only implemented for MPS in the singlet-embedding\cite{sharma2012spin} format.

(ii) We can use this algorithm to extract the  \( T_3 \) and \( T_4 \) coupled cluster amplitudes, required in  externally corrected CCSD\cite{magoulas2021externally} where the DMRG wavefunction is the external source.\cite{lee2021externally} For this purpose, we use a much smaller threshold for the CI coefficient, and an additional restriction on the excitation order (up to quadruples) with respect to the Fermi vacuum to preserve the polynomial complexity of the overall algorithm.

(iii) The same extraction algorithm can also be utilized in a tailored CCSD\cite{kinoshita2005coupled,hino2006tailored} code using the DMRG wavefunction as the external source for extracting the \( T_1 \) and \( T_2 \) coupled cluster amplitudes in the active space.\cite{veis2016coupled,faulstich2019numerical}

\subsection{MPS transformations}

In \textsc{block2} the MPS may be converted between different representations suitable for different tasks.
There are some limitations on the operations supported in different MPS representations. Below, we discuss examples of such MPS transformations.

\subsubsection{Singlet embedding}

A spin-adapted MPS with non-zero total spin can be represented in either the singlet embedding (SE)\cite{mcculloch2002non,sharma2012spin,li2017spin} format (with additional non-interacting fictitious spins added to the end of the DMRG lattice to create a singlet) or the non-singlet-embedding (NSE) format. 

For MPS expressed in the SE format, there is no accuracy loss when changing the canonical form between sites, so the DMRG optimization process becomes more stable. However, the SE MPS form makes it difficult to evaluate certain quantities, such as the one-particle triplet transition density matrix (1-TTDM),\cite{li2017spin}  required in the 2-step treatment of the SOC effect.\cite{sayfutyarova2016state} On the other hand, we do not support the 
computation of CSF coefficients from NSE MPS. To overcome these limitations, we have implemented the exact transformation from SE to NSE MPS, and from NSE to SE MPS. The user can perform the optimization of MPS using either the SE or NSE format, and then transform the MPS to the appropriate format before computing the 1-TTDM or extracting CSF coefficients.

\subsubsection{Orbital rotations}

To improve the representation power of a given bond dimension MPS in DMRG, it is sometimes desirable to rotate the orbital basis.\cite{mitrushchenkov2012importance} In \textsc{block2} we use an exponential parametrization of the orbital rotation matrix,\cite{shepard2015representation} and implement the orbital rotation as a time evolution.\cite{feiguin2005time,ronca2017time} As an example usage of this transformation, one can perform DMRG optimization using a set of localized orbitals, and then compute the determinant coefficients defined in a different set of orbitals after an MPS orbital rotation.

As the MPS entanglement structure in two sets of orbitals can be significantly different, often the MPS orbital rotation cannot be performed exactly. When a fixed bond dimension and time step is used for the transformation, the resulting MPS can sometimes be highly inaccurate. We provide a few additional options to improve the accuracy.

(i) When we want the orbital transformation to approximately preserve the orbital ordering, we use the Kuhn-Munkres algorithm\cite{kuhn1955hungarian,fukuda1994finding} to first match the old and new orbitals.

(ii) We can remove some artificial rotations in the logarithm of the rotation matrix by forcing all leading principal minors of the rotation matrix to be non-negative.

\subsubsection{Symmetry mappings}

As described earlier, \textsc{block2} supports many kinds of symmetries in the DMRG calculations. We also provide the ability to transform MPS from a higher to a lower symmetry group. For example:

(i) We can transform spin-adapted (SA) MPS\cite{sharma2012spin,wouters2014chemps2,keller2016spin} to non-spin-adapted (NSA) MPS. This can be done exactly, and the bond dimension of the transformed NSA MPS will generally be higher than that of the SA MPS.\cite{sharma2012spin} Since the DMRG optimization of the SA MPS is often much cheaper, this can be useful when the analysis has to be performed on the NSA MPS (for example, when information on the determinant expansion is desired). When the SA MPS is not a singlet, the user can additionally set the desired projected spin of the transformed NSA MPS, and the transformed MPS will have both well-defined total and projected spin. 

(ii) We can transform from an MPS expressed in one point group to a subgroup. To perform such a transformation, the user must provide a mapping between the irreducible representations in the two groups. In \textsc{block2} we implement this type of MPS symmetry mapping using a fitting approach, where a small fitting error can be expected.

\subsection{DMRG for dynamical and time-dependent quantities}

\textsc{block2} supports a full suite of  algorithms to compute dynamical quantities, such as Green's functions, as well as to perform time-dependent state propagation.

\subsubsection{Dynamical DMRG}

Dynamical DMRG (DDMRG) is the original extension of the standard DMRG algorithm to compute dynamical quantities, such as Green's functions and other spectroscopic quantities~\cite{jeckelmann2002dynamical,ramasesha1997low,dorando2009analytic,ronca2017time,lee2023ab} [we will use the electron removal (IP) part of one-particle Green's function in the examples below]. In \textsc{block2}, we provide an improved version of dynamical DMRG called \( \mathrm{DDMRG}^{++}\),  introduced in \lit{ronca2017time}. We provide the following options for this method.

(i) We can solve the response equations to find the correction vector as an MPS, and the IP one-particle Green's function is obtained as the expectation value
\begin{align}
    G^-_{ij}(\omega) :=&\ \langle \Psi_0 | a_j^\dagger | \Psi'_i(\omega) \rangle \\
    [\hat{H}_0 - E_0 +\omega -\mathrm{i} \eta ] |\Psi'_i(\omega) \rangle =&\ a_i |\Psi_0 \rangle
    \label{eq:ddmrg-cpx}
\end{align}
where \( |\Psi_0 \rangle \) and \( E_0 \) are the (assumed real) ground state MPS and energy, \( |\Psi'_i(\omega) \rangle \) is the correction vector MPS, \( G^-_{ij}(\omega) \) is the IP part of one-particle Green's function, and \( \omega \) and \( \eta \) are the real frequency and broadening factor, respectively. In general, \( |\Psi'_i(\omega) \rangle \) will be complex, and we solve this equation in the complex domain using the \( \mathrm{DDMRG}^{++}\) sweep algorithm.

(ii) Alternatively, we  write the correction vector as \( |\Psi'_i(\omega) \rangle := |X_i(\omega) \rangle + \mathrm{i} |Y_i(\omega) \rangle \) and \autoref{eq:ddmrg-cpx} becomes\cite{jeckelmann2002dynamical,ronca2017time}
\begin{equation}
    \big[ \big(\hat{H}_0 - E_0 +\omega\big)^2 + \eta^2 \big]
    |Y_i(\omega) \rangle = \eta a_i |\Psi_0 \rangle
    \label{eq:ddmrg-real}
\end{equation}
In \textsc{block2} we also use \( \mathrm{DDMRG}^{++}\) to solve the above equation in the real domain.

(iii) For very small \( \eta \), both \autoref{eq:ddmrg-cpx} and \autoref{eq:ddmrg-real} will become ill-conditioned for some \( \omega \), and the required number of CG or GCROT\cite{hicken2010simplified} iterations at each site during the sweep can become large.\cite{wathen2015preconditioning}  To accelerate the computation of the DMRG Green's function, we  support a multi-grid frequency strategy. At the coarse-grained level, we use a large \( \eta_0 \) and a coarse grid of \( \omega_0 \) for the DDMRG sweeps. At the fine-grained level, we solve \autoref{eq:ddmrg-cpx} or \autoref{eq:ddmrg-real} using a smaller \( \eta \) only in the middle site in the sweep, for a few \( \omega \) near the \( \omega_0 \) value. This method assumes that the renormalized basis in the correction vector MPS generated at \( \omega_0 \) with a large broadening can be reused for adjacent frequencies. Using this strategy, we can often achieve a good balance between accuracy and efficiency.

\subsubsection{Chebyshev DMRG}

To avoid solving the ill-conditioned response equation in DDMRG, we have also implemented the Chebyshev DMRG approach,\cite{holzner2011chebyshev,braun2014numerical,xie2018reorthonormalization,jiang2021chebyshev} based on the Chebyshev expansion of the resolvent and the MPS representation of the Chebyshev vectors.

\subsubsection{Time-dependent DMRG}

\label{sec:tddmrg}

\textsc{block2} implements time-dependent DMRG (td-DMRG) for true non-equilibrium dynamics, as well as to simulate linear spectra (via the Fourier transform of the autocorrelation function).

(i) We support both imaginary time evolution (ITE) and real time evolution (RTE). When the Hamiltonian is real (and the initial state is real) ITE can be performed using only real arithmetic, while RTE can be carried out in hybrid real-and-complex mode, where the MPO and rotation matrices in the MPS are represented as real objects, and the MPS tensor at the canonical center is complex. Alternatively, one can perform time evolution where both MPO and MPS are complex and the time step can also be a complex number. 

(ii) For the sweep algorithm used in td-DMRG, we implemented both the time-step targeting (TST) approach\cite{feiguin2005time} and the time-dependent variational principle\cite{dorando2009analytic,haegeman2011time,kinder2014analytic,nakatani2014linear,haegeman2016unifying} approach. For the TST approach, we support both the td-DMRG and td-\(\mathrm{DMRG}^{++} \)\cite{ronca2017time} variants. The Hamiltonian involved in these algorithms can be Hermitian or non-Hermitian (the latter appears when time evolution is used to perform orbital rotation). 

\begin{figure}[!htbp]
  \includegraphics[width=\columnwidth]{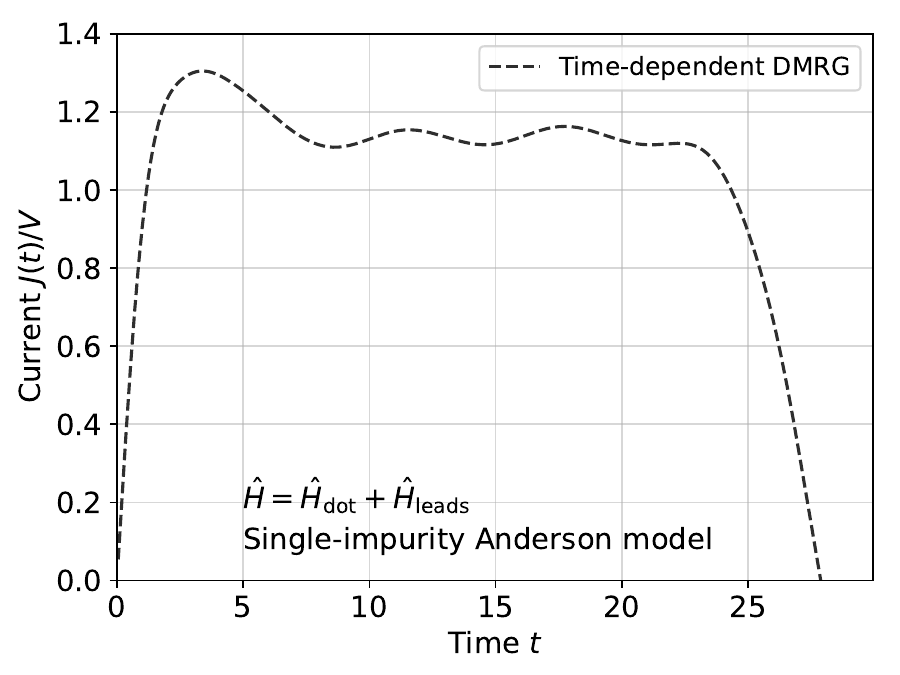}
  \caption{Current as a function of time for the single impurity Anderson model with \( U = 0.5t, t' = 0.3535t, V = 2U, V_{\mathrm{g}} = -0.5U\) and 48 sites, simulated using time-dependent DMRG with the time-dependent variational principle approach, time step \( \Delta t = 0.025 \), and MPS bond dimension \( M = 1000 \). (The curve is chosen to illustrate dynamics rather than full convergence with respect to all parameters.)  Model parameters are taken from \lit{heidrich2009real}.}
  \label{fig:tddmrg}
\end{figure}

\autoref{fig:tddmrg} shows a numerical example using time-dependent DMRG.

\subsection{Finite-temperature DMRG}

\textsc{block2} supports several finite-temperature DMRG algorithms. 

\subsubsection{Ancilla approach}

In the ancilla approach for finite temperature,\cite{feiguin2005finite} we perform imaginary time evolution of a purified thermal state in an enlarged Hilbert space. Within the context of DMRG, we represent the purified state as an MPS, where ancilla sites are added  as a bath. In \textsc{block2} we provide subroutines to create the initial finite-temperature MPS at inverse temperature \( \beta = 0 \) and the finite-temperature MPO, which is a normal MPO decorated with identity operators at the ancilla sites. We support the evaluation of free energies and PDMs of the finite-temperature MPS by tracing out the ancilla sites. The finite-temperature variants of dynamical DMRG\cite{jiang2020finite} and time-dependent DMRG\cite{ren2018time,peng2021conservation} are also available to compute dynamical quantities.

\begin{figure}[!htbp]
  \includegraphics[width=\columnwidth]{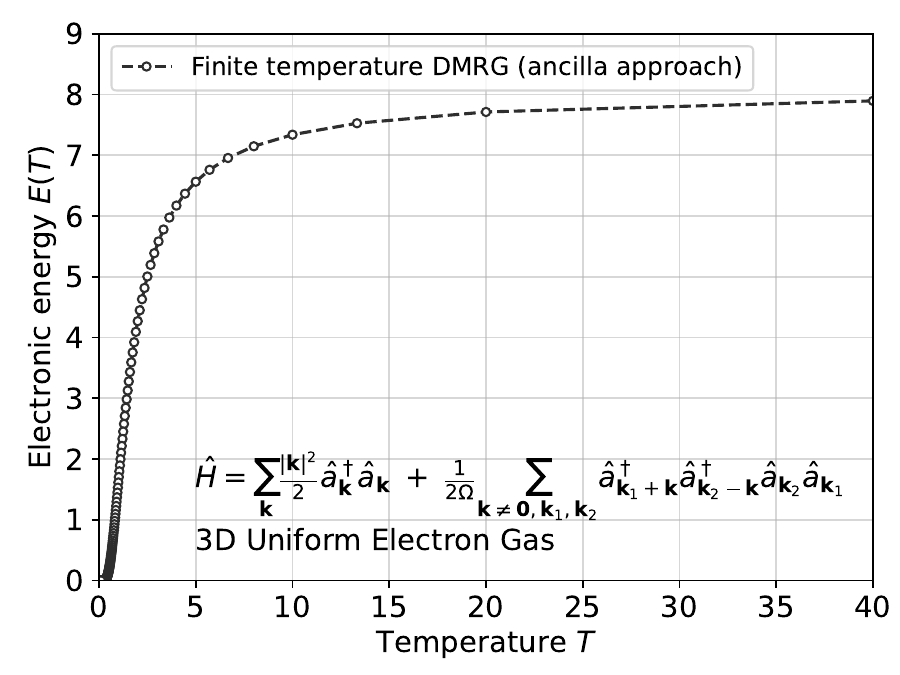}
  \caption{Electronic energy at different temperatures for the three dimensional uniform electron gas model\cite{loos2016uniform} with density \(r_s = 4\), box length \( L=10.24 \), and plane wave basis set kinetic energy cutoff \( E_{\mathrm{cut}} = 0.6199 \), simulated using finite-temperature DMRG with the ancilla approach\cite{feiguin2005finite} and RK4 for imaginary time evolution,\cite{feiguin2005time} at chemical potential \( \mu = -2 \).}
  \label{fig:ftdmrg}
\end{figure}

\autoref{fig:ftdmrg} shows a numerical example of the finite-temperature DMRG using the ancilla approach.

\subsubsection{Sum-over-states approach}

To compute properties at very low temperatures (namely, when \( \beta \) is large), the ancilla approach can be inefficient since a large number of time steps is required and this can also generate large entanglement between the physical and ancillary degrees of freedom. We instead  perform time-independent DMRG to find a few low-lying eigenstates, and the finite-temperature effects are then computed through a partition function sum over these states. Since the low-lying states can have different particle numbers and spins, we can either represent them using a single particle-number-non-conserving state-averaged MPS, so that the required number of states in each symmetry sector can be determined automatically, or by using multiple state-averaged MPSs to cover all symmetry sectors. The latter scheme may be more efficient, but requires the knowledge of the number of low-energy states in each symmetry sector.

\subsection{DMRG and dynamical correlation}

When simulating realistic chemical problems, the number of correlated orbitals can quickly exceed the capabilities of the standard DMRG algorithm for quantum chemistry. Thus it is necessary to combine DMRG with other ideas to treat both static and dynamic correlation effects.\cite{szalay2012multiconfiguration,yanai2015density,cheng2022post,yanai2010multireference,kurashige2011second,guo2016n,saitow2013multireference,sharma2015multireference,luo2018externally,sharma2019density,beran2021density,khokhlov2021toward} In the following we list a few different dynamical correlation treatments implemented in \textsc{block2}.

\subsubsection{Perturbative DMRG}

In perturbative DMRG,\cite{guo2018perturbative} the energy of a small bond dimension variational DMRG calculation is corrected using second-order perturbation theory, where the Hamiltonian is partitioned similarly to the Epstein-Nesbet partitioning. In \textsc{block2}, we include two implementations of this method (for both non-spin-adapted DMRG and spin-adapted DMRG).

(i) In the deterministic perturbative DMRG approach,\cite{guo2018perturbative} the first-order perturbative wavefunction is represented as an MPS, which is optimized by solving the response equation using dynamical DMRG (without frequency).

(ii) In the stochastic perturbative DMRG approach,\cite{guo2018communication} we first compress the first-order wavefunction to a low bond dimension MPS, then the second-order energy correction is evaluated by determinant or CSF sampling.

\subsubsection{Uncontracted dynamical correlation theories}

It is natural to add dynamical correlation to DMRG in the framework of multireference theories, where the DMRG-CASSCF solution is used as a reference state, and low-order excitations from this space are used for the dynamic correlation treatment. We can represent the wavefunction ansatz (including the active, core, and virtual sub-spaces) as an MPS that approximates an uncontracted correlation ansatz.\cite{sharma2014communication,sharma2016quasi} As introduced in a recent report,\cite{larsson2022matrix} we provide a few different implementations to represent and optimize MPS for uncontracted dynamical correlation theories.

(i) We can construct a conventional MPS that describes all the orbital spaces, but restrict the quantum numbers to limit the excitation order out of the active space.\cite{larsson2022matrix} This approach can be trivially supported in \textsc{block2}, so it is automatically compatible with many other features. However, its efficiency can be low. When the excitation order is restricted to singles, we provide an accelerated solver that bypasses the sweep iterations over the external orbitals.\cite{sharma2017combining}

(ii) For improved efficiency, we can introduce two large sites in the MPS to describe the core and virtual spaces so that operations on these orbitals can be handled specially.\cite{larsson2022matrix} In \textsc{block2}, we support several different treatments of the large site. The CSF and determinant based large-site implementation provides the highest efficiency, for the spin-adapted and non-spin-adapted cases, respectively. We also support building a large-site MPO tensor by contracting normal MPO tensors, which is less efficient.

Using the large-site scheme, we can optimize the MPS energy for different multireference theories (with large active spaces and many core orbitals),\cite{larsson2022matrix} including arbitrary-order multi-reference configuration interaction (MRCI)\cite{larsson2022matrix,barcza2022toward} and size-extensivity corrections,\cite{szalay2012multiconfiguration} arbitrary-order averaged quadratic coupled cluster (AQCC),\cite{szalay1993multi} arbitrary-order coupled pair functional (ACPF),\cite{gdanitz1988averaged} and arbitrary-order NEVPT\cite{angeli2001introduction,angeli2002n,angeli2007new} and restraining the excitation degree multireference perturbation theory (MRREPT).\cite{fink2006two,fink2009multi,sharma2015multireference}Our flexible implementation further allows for the inclusion of arbitrary determinants/CSFs in a fashion similar to selected configuration interaction.

\subsubsection{Internally contracted dynamical correlations}

Internally contracted dynamical correlation theories form a second common framework for multireference correlation. When the active space indices in the excitation operator are also contracted, the corresponding theories are called strongly-contracted.\cite{guo2016n,roemelt2016projected,sokolov2017time} Currently we provide the spin-free implementations of fully internally contracted MRCI singles and doubles (FIC-MRCISD),\cite{saitow2013multireference,saitow2015fully} internally contracted NEVPT2,\cite{sharma2017combining} and strongly contracted NEVPT2\cite{guo2016n} using automatic symbolic expression derivation and DMRG \( N \)-PDMs implemented in \textsc{block2}. Other internally contracted theories are supported via the interface between \textsc{block2} and some external packages, which will be discussed in the next section.

\begin{figure}[!htbp]
  \includegraphics[width=\columnwidth]{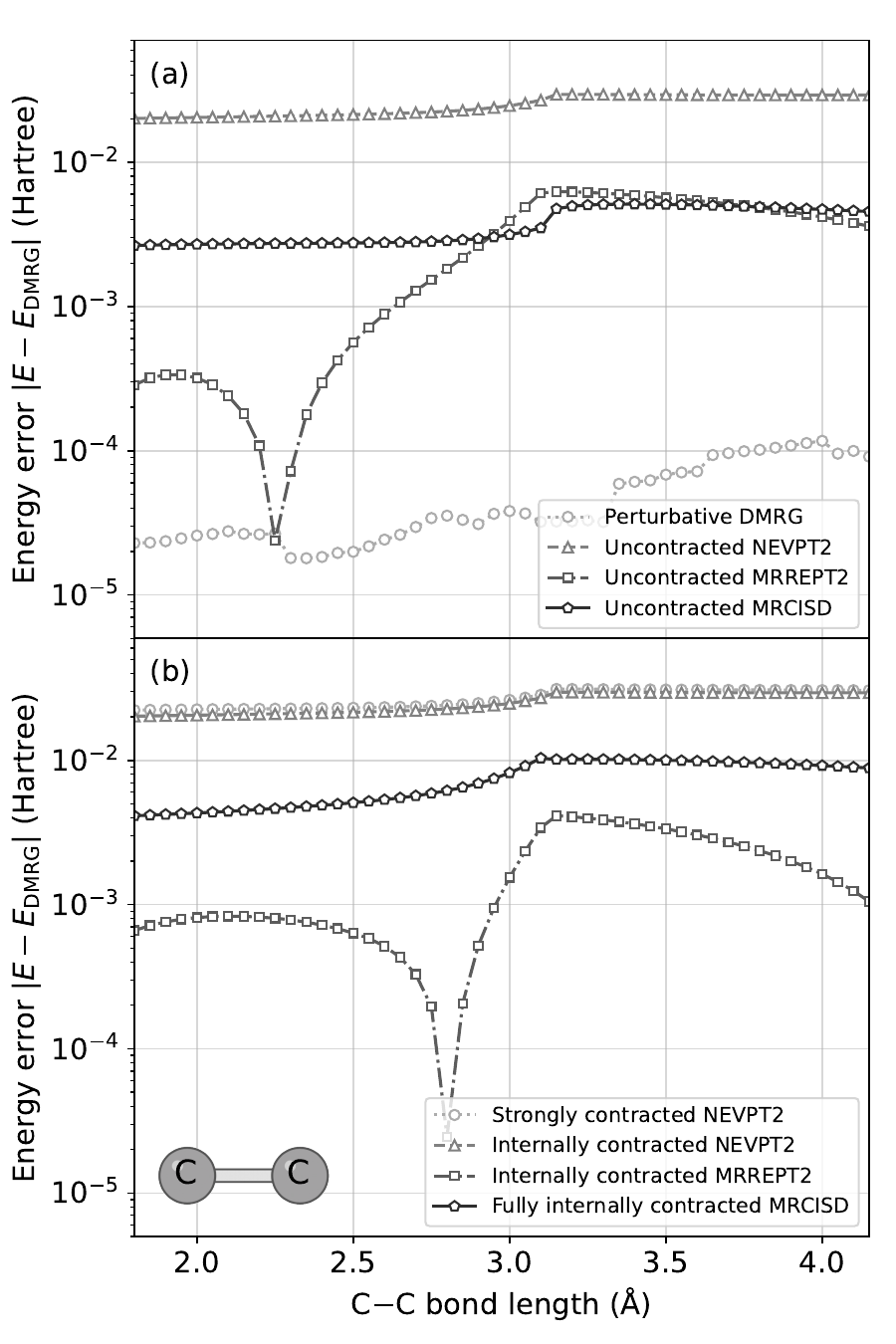}
  \caption{A comparison between absolute energies computed using DMRG (with MPS bond dimension \( M = 3000 \)) and (a) perturbative DMRG and uncontracted and (b) internally contracted dynamic correlation multi-reference theories implemented in \textsc{block2}, for carbon dimer in the cc-pVDZ basis set at different bond lengths. DMRG-CASSCF with a active space (8o, 8e) is performed to generate the reference state for MRCISD, NEVPT2, and MRREPT2. The reference state for perturbative DMRG is an MPS with \( M = 200 \).}
  \label{fig:mr}
\end{figure}

\autoref{fig:mr} shows a numerical example of the performance of theories for DMRG with dynamical correlations.

\section{Interfaces and Extensions}

To extend the range of application of  DMRG algorithms, we have designed \textsc{block2} to be interoperable with other libraries. In this section, we list some of the available basic interfaces of \textsc{block2}, as well as the interfaces between  \textsc{block2} and external packages. 

\subsection{Basic interfaces to \textsc{block2}}

The source code of \textsc{block2} is mainly written in C++11, as a header-only library. We use templates extensively for a concise treatment of different symmetry classes and floating point number types. The code can be compiled using GNU gcc, clang, Apple clang, Intel icc, or the msvc compilers, and supports the Linux, MacOS, and Windows platforms. For MacOS we support both the x86 and arm64 instruction architectures. Currently, development and testing focus on the Linux (x86) and MacOS (x86) platforms. We currently provide the following basic interfaces to \textsc{block2}.

(i) A Python library interface to \texttt{block2} and \texttt{pyblock2} (not to be confused with \texttt{pyblock3}). The former exposes almost all C++ classes and functions to Python using \textsc{pybind11}.\cite{jakob2017pybind11} The latter provides some helper and driver classes written in pure Python, for users to easily customize their workflow.

(ii) The Python script \texttt{block2main} for parsing input files and running DMRG. This provides a simple interface which should be familiar to users of conventional quantum chemistry packages. The input file format is also compatible with that of \textsc{StackBlock},\cite{stackblock} so existing applications that call \textsc{StackBlock} as an executable can be directly used with \textsc{block2}. A list of the supported keywords is given in the documentation.

(iii) A C++ executable \texttt{block2} to parse input files and run DMRG. We have found this interface to be useful when performing very large scale distributed parallel computations, where the Python components can induce significant overheads.

(iv) The C++ library \texttt{libblock2.so}, which can be used together with other packages written in C++. In principle, one can also simply include \textsc{block2} header files in a C++ project without linking to the library. However, in practice this may introduce unnecessary extra compilation time.

Besides providing freely downloadable source code,\cite{block2} we also provide precompiled packages for the Python interfaces, which can be installed using \texttt{pip install block2} [without the Message Passing Interface (MPI) library dependence] or \texttt{pip install block2-mpi} (with MPI dependence).

\subsection{Extension interfaces to external packages}

We provide the following interfaces to external packages to extend the functionality of \textsc{block2}.

(i) \textsc{PySCF}.\cite{sun2018pyscf,sun2020recent} The orbitals and integrals required for running DMRG can be computed using mean-field methods implemented in \textsc{PySCF}. One can also perform DMRG-CASSCF (energy and gradients),\cite{ghosh2008orbital} DMRG SC-NEVPT2 (using the exact 4-PDM\cite{guo2016n} or MPS compression\cite{sokolov2017time}) with \textsc{PySCF} and \textsc{block2} (this partially reuses the existing \textsc{PySCF} interface to \textsc{StackBlock}, which gives an alternative implementation of SC-NEVPT2 to the native implementation in \textsc{block2}).
Using the Python interface to the \textsc{block2} 
symbolic engine and some of the \textsc{PySCF} infrastructure, one can also automatically derive and implement a variety of coupled cluster theories.

(ii) \textsc{LibDMET}.\cite{cui2019efficient} One can use ground state or finite temperature DMRG as the impurity solver\cite{sun2020finite} for \emph{ab initio} Density Matrix Embedding Theory (DMET) through the interface between \textsc{block2} and \textsc{LibDMET}.\cite{cui2022systematic} In particular, the \emph{ab initio} DMRG with general spin orbitals was implemented as an impurity solver for the superconducting Hamiltonian.\cite{cui2023ab}

(iii) \textsc{fcdmft}.\cite{zhu2019efficient,zhu2021ab} One can use dynamical DMRG or DMRG uncontracted MRCI as the impurity solver for \emph{ab initio} full cell Dynamical Mean-Field Theory (DMFT) using \textsc{block2} with \textsc{fcdmft}.

(iv) \textsc{StackBlock}.\cite{stackblock,sharma2012spin} We provide scripts to transform MPSs (in both directions) between \textsc{StackBlock} and \textsc{block2}. This is primarily useful for benchmarking purposes.

(v) \textsc{PyBlock3}.\cite{pyblock3} We provide scripts to transform MPSs and MPOs in both directions between \textsc{PyBlock3} and \textsc{block2}.

(vi) \textsc{OpenMolcas}.\cite{aquilante2020modern} We can perform DMRG-CASSCF and DMRG-CASPT2\cite{kurashige2011second,liu2013multireference,wouters2016dmrg,nakatani2017density} (with or without cumulant approximations\cite{kurashige2014complete}) using \textsc{OpenMolcas} and \textsc{block2}. This is implemented using a small modification to the existing interface between \textsc{OpenMolcas} and \textsc{StackBlock}.\cite{phung2016cumulant}

(vii) \textsc{Forte}.\cite{li2019multireference} One can perform DMRG-CASCI, DMRG-CASSCF, and DMRG-Driven-Similarity-Renormalization-Group (DSRG)\cite{khokhlov2021toward} with the interface between \textsc{Forte} and \textsc{block2}.

(viii) \textsc{Qiskit}.\cite{qiskit} Through the Python interface in \textsc{block2} one can perform fermionic or spin DMRG for Hamiltonian expressions generated in \textsc{Qiskit}. For example, one can perform fermionic DMRG from the \textsc{Qiskit} second quantized Hamiltonian expression, or DMRG for Hamiltonians expressed as linear combinations of arbitrary-length Pauli strings.\cite{mishmash2023hierarchical}

As suggested by these examples, it is also simple to extend \textsc{block2} interfaces to work with many other packages.

\section{Conclusions}

After more than three years of development, the robustness, utility, and comprehensiveness of the \textsc{block2} framework has been examined and tested across a great variety of projects. In our experience, the range of implementations and algorithms provided by \textsc{block2} has allowed DMRG to be fruitfully applied for many different purposes, ranging from benchmarking other methods, to obtaining high-accuracy definitive results in \emph{ab initio} and model simulations. Moving forwards, we will continue to 
develop 
\textsc{block2} as an open framework for DMRG algorithms and to further extend its integration with other packages. 
We hope that \textsc{block2} will serve as platform to 
extend DMRG to new domains and to generate insights into new scientific areas.

\begin{acknowledgments}
H.Z. thanks Sandeep Sharma for helpful discussions.
This framework was produced over several years with contributions from multiple individuals.
Work by H.Z. was supported by the US National Science Foundation, under Award No. CHE-2102505.
Work by H.R.L. and T.Z. was supported by the Air Force Office of Scientific Research, under Award FA9550-18-1-0095. 
H.R.L.~acknowledges support from a postdoctoral fellowship from the German Research Foundation (DFG) via grant LA 4442/1-1 during the first part of this work. 
Work by S.L. (this material) was supported by the US Department of Energy, Office of Science, National Quantum Information Science Research Centers, Quantum Systems Accelerator.
Work by Z.C. was supported by the US Department of Energy, Office of Science, under Award No. DE-SC0018140.
Work by C.S. was supported by the US Department of Energy, Office of Science, under Grant No. DE-SC0019374.
Work by L.P. was supported by the US Department of Energy, Office of Science, via the M2QM EFRC under Grant No. DE-SC0019330.
Work by R.P. was supported by the US National Science Foundation, under Award No. CHE-2102505.
Work by K.L. was supported by the US Department of Energy, Office of Science, Basic Energy Sciences and  Office of Advanced Scientific Computing Research, Scientific Discovery through Advanced Computing (SciDAC) program, under Award No. DE-SC0022088.
Work by J.T. was supported by the US Department of Energy, Office of Science, under Award No. DE-SC0018140.
J.T. acknowledges funding through a postdoctoral research fellowship from the Deutsche Forschungsgemeinschaft (DFG, German Research Foundation)-495279997.
Work by J.Y. was supported by the US Department of Energy, Office of Science, via the M2QM EFRC under Grant No. DE-SC0019330. The computations presented in this work were conducted at the Resnick High Performance Computing Center, a facility supported by the Resnick Sustainability Institute at the California Institute of Technology.
\end{acknowledgments}

\section*{Data Availability}
The reference input and output files for producing data in the numerical examples can be found in the GitHub repository \url{https://github.com/hczhai/block2-example-data}. The reference software version is \textsc{block2} 0.5.2.

\bibliography{main}

\end{document}